\documentclass[useAMS,usenatbib]{mn2e}
\usepackage{mn2e-breakabs}
\usepackage{graphicx}
\usepackage{times}
\def\la{\mathrel{\mathpalette\fun <}}
\def\ga{\mathrel{\mathpalette\fun >}}
\def\fun#1#2{\lower3.6pt\vbox{\baselineskip0pt\lineskip.9pt
        \ialign{$\mathsurround=0pt#1\hfill##\hfil$\crcr#2\crcr\sim\crcr}}}

\def\bfk{\mbox{\bf k}}
\def\bfr{\mbox{\bf r}}
\def\etal{{\frenchspacing\it et al.}}

\def\rd{{\rm d}}

\newcommand{\be}{\begin{equation}}
\newcommand{\ee}{\end{equation}}
\newcommand{\ba}{\begin{eqnarray}}
\newcommand{\ea}{\end{eqnarray}}
\newcommand{\simgt}{\,\hbox{\lower0.6ex\hbox{$\sim$}\llap{\raise0.6ex\hbox{$>$}}}\,}
\newcommand{\simlt}{\,\hbox{\lower0.6ex\hbox{$\sim$}\llap{\raise0.6ex\hbox{$<$}}}\,}

\begin{document}

\title[Galaxy clustering]
{Robust constraints on dark energy and gravity \\from galaxy clustering data}

\author[Yun Wang]{
  \parbox{\textwidth}{
    Yun Wang\thanks{E-mail: wang@nhn.ou.edu}}
  \vspace*{4pt} \\
  Homer L. Dodge Department of Physics \& Astronomy, Univ. of Oklahoma,
                 440 W Brooks St., Norman, OK 73019, U.S.A.}

\date{\today} 

\maketitle

\begin{abstract}
Galaxy clustering data provide a powerful probe of dark energy.
We examine how the constraints on the scaled expansion history of the universe, $x_h(z)=H(z)s$ 
(with $s$ denoting the sound horizon at the drag epoch), and
the scaled angular diameter distance, $x_d(z)=D_A(z)/s$, depend on the methods used to analyze
the galaxy clustering data. We find that using the observed galaxy power spectrum, $P_g^{obs}(k)$,
$x_h(z)$ and $x_d(z)$ are measured more accurately and are significantly less correlated
with each other, compared to using only the information from the baryon acoustic oscillations
(BAO) in $P_g^{obs}(k)$. Using the $\{x_h(z), x_d(z)\}$ from $P_g^{obs}(k)$ gives a DETF dark energy
FoM approximately a factor of two larger than using the $\{x_h(z), x_d(z)\}$ from BAO only;
this provides a robust conservative method to go beyond BAO only in extracting 
dark energy information from galaxy clustering data.

We find that a Stage IV galaxy redshift survey, with $0.7<z<2$ over 15,000 (deg)$^2$,
can measure $\{x_h(z), x_d(z), f_g(z)G(z)\tilde{P}_0^{1/2}/s^4\}$ with high precision
(where $f_g(z)$ and $G(z)$ are the linear growth rate and factor of large scale structure respectively, 
and $\tilde{P}_0$ is the dimensionless normalization of $P_g^{obs}(k)$), when redshift-space distortion information
is included. The measurement of $f_g(z)G(z)\tilde{P}_0^{1/2}/s^4$ provides a powerful test of gravity,
and significantly boosts the dark energy FoM when general relativity is assumed.

\end{abstract}


\section{Introduction}  \label{sec:intro}

The cosmic acceleration (i.e., dark energy) was discovered in 1998 \citep{Riess98,Perl99}, and 
we are still in the dark about the nature of this mystery. We can hope to measure
both the cosmic expansion history and the cosmic large scale structure growth history accurately 
and precisely with galaxy clustering (see, e.g., \cite{Guzzo08,Wang08a}) and weak lensing 
(see, e.g., \cite{Knox06,Zhang07,Heavens09})
data from a space mission such as Euclid \citep{RB}\footnote{http://www.euclid-ec.org}, 
and differentiate the two possible explanations for the observed cosmic acceleration: 
a new energy component, or a modification of Einstein's theory of gravity.\footnote{Clusters
of galaxies provide a complementary probe of dark energy, see, e.g., 
\cite{Majumdar04,Manera06,Mota08,Sartoris11}.}

Galaxy clustering has long been used as a cosmological probe (see, e.g., \cite{Hamilton98}). 
At present, the largest data set comes from the SDSS III Baryon Oscillation Spectroscopic Survey (BOSS),
see \cite{Anderson12} and \cite{Reid12}.

Here we explore different analysis techniques for galaxy clustering data, in order to obtain robust 
constraints on dark energy and general relativity. It is important to study how the dark energy and
gravity constraints depend on the assumptions we make about the information that can be extracted from galaxy 
redshift survey data.

We present our methods in Sec.2, results in Sec.3, and summarize in Sec.4.

\section{Methodology} \label{sec:methods}

\subsection{The Fisher Matrix Formalism}

We use the Fisher matrix formalism to study the parameter estimation using
galaxy clustering data \citep{Tegmark97,SE03}, based on the approach developed in \cite{Wang06,Wang08a,Wang10,Wang_etal10}.
In the limit where the length scale corresponding to the survey volume is 
much larger than the scale of any features in the observed galaxy power spectrum
$P_g(k)$, we can assume that the likelihood function for the band powers of a galaxy 
redshift survey is Gaussian \citep{FKP}, with a measurement error
in $\ln P(\bfk)$ that is proportional to $[V_{eff}(\bfk)]^{-1/2}$, 
with the effective volume of the survey defined as
\ba
V_{eff}(k,\mu)&\equiv& \int d\bfr^3 \left[ \frac{n(\bfr) P_g(k,\mu)}
{ n(\bfr) P_g(k,\mu)+1} \right]^2\nonumber\\
&=&\left[ \frac{ n P_g(k,\mu)}{n P_g(k,\mu)+1} \right]^2 V_{survey},
\ea
where the comoving number density $n$ is assumed to only depend on
the redshift (and constant in each redshift slice) for simplicity
in the last part of the equation.

In order to propagate the measurement error in $\ln P_g(\bfk)$ 
into measurement errors for the parameters $p_i$, we use
the Fisher matrix \citep{Tegmark97}
\be
F_{ij}= \int_{k_{min}}^{k_{max}}
\frac{\partial\ln P_g(\bfk)}{\partial p_i}
\frac{\partial\ln P_g(\bfk)}{\partial p_j}\,
V_{eff}(\bfk)\, \frac{d \bfk^3}{2\, (2\pi)^3}
\label{eq:Fisher_full}
\ee
where $p_i$ are the parameters to be estimated from data, and 
the derivatives are evaluated at parameter values of the
fiducial model. Note that the Fisher matrix $F_{ij}$ is the 
inverse of the covariance matrix of the parameters $p_i$ if 
the $p_i$ are Gaussian distributed.

The observed galaxy power spectrum can be reconstructed using a particular reference 
cosmology, including the effects of bias and redshift-space distortions 
\citep{SE03}:
\ba
\label{eq:P(k)b}
P_{g}^{obs}(k^{ref}_{\perp},k^{ref}_{\parallel}) &=&
\frac{\left[D_A(z)^{ref}\right]^2  H(z)}{\left[D_A(z)\right]^2 H(z)^{ref}}
\, b^2 \left( 1+\beta\, \mu^2 \right)^2 \cdot\nonumber\\
&& \,\,\,\cdot [G(z)]^2 P_{m}(k)_{z=0}+ P_{shot},
\ea
where $H(z)=\dot{a}/{a}$ (with $a$ denoting the cosmic scale factor) is
the Hubble parameter, and $D_A(z)=r(z)/(1+z)$ is the angular diameter distance at $z$,
with the comoving distance $r(z)$ given by
\be
\label{eq:r(z)}
 r(z)=c\, |\Omega_k|^{-1/2} {\rm sinn}\left[|\Omega_k|^{1/2}\, 
\int_0^z\frac{dz'}{H(z')}\right],
\ee
where ${\rm sinn}(x)=\sin(x)$, $x$, $\sinh(x)$ for 
$\Omega_k<0$, $\Omega_k=0$, and $\Omega_k>0$ respectively.
The bias between galaxy and matter distributions is denoted by
$b(z)$. The linear redshift-space distortion parameter
$\beta(z)=f_g(z)/b(z)$ \citep{Kaiser87}, where $f_g(z)$ is the linear growth rate;
it is related to the linear growth factor $G(z)$ 
(normalized such that $G(0)=1$) as follows 
\be
f_g(z)=\frac{\mbox{d}\ln G(z)}{\mbox{d}\ln a}.
\ee

Note that $\mu = \bfk \cdot \hat{\bfr}/k$, with $\hat{\bfr}$ denoting the unit
vector along the line of sight; $\bfk$ is the wavevector with $|\bfk|=k$.
Hence $\mu^2=k^2_{\parallel}/k^2=k^2_{\parallel}/(k^2_{\perp}+k^2_{\parallel})$,
where
\ba
k_\parallel &=& \bfk \cdot \hat{\bfr}=k\mu  \nonumber\\
k_\perp &= &\sqrt{k^2-k^2_\parallel}=k\sqrt{1-\mu^2}.
\ea
The values in the reference cosmology are denoted by the subscript ``ref'',
while those in the true cosmology have no subscript.
Note that 
\be
k^{ref}_{\perp}=k_\perp D_A(z)/D_A(z)^{ref}, \hskip 0.5cm
k^{ref}_{\parallel}=k_\parallel H(z)^{ref}/H(z).
\ee

Eq.(\ref{eq:P(k)b}) characterizes the dependence of the observed galaxy power
spectrum on $H(z)$ and $D_A(z)$ due to BAO, as well as 
the sensitivity of a galaxy redshift survey to the linear redshift-space 
distortion parameter $\beta$ \citep{Kaiser87}.
The linear matter power spectrum at $z=0$ is given by
\be
P_m(k)_{z=0}=P_0\, k^{n_s} T^2(k),
\label{eq:P(k|z=0)}
\ee
where $T(k)$ denotes the matter transfer function.

The measurement of $f_g(z)$ given $\beta(z)=f_g(z)/b(z)$ requires an additional
measurement of the bias $b(z)$, which could be obtained from the
galaxy bispectrum \citep{Matarrese97,Verde02}. 
Alternatively, we can rewrite Eq.(\ref{eq:P(k)b}) as
\ba
\label{eq:P(k)fg_scale1}
&&\overline{P_g^{obs}(k^{ref}_{\perp},k^{ref}_{\parallel})}\nonumber\\
&\equiv &P_g^{obs}(k^{ref}_{\perp},k^{ref}_{\parallel})/(h^{-1}\mbox{Mpc})^3 \nonumber\\
&=&\frac{\left[D_A(z)^{ref}\right]^2  H(z)}{\left[D_A(z)\right]^2 H(z)^{ref}}
 \left(\frac{k}{\mbox{Mpc}^{-1}}\right)^{n_s} T^2(k) \cdot \nonumber\\
& & \cdot  \left[\sigma_{g}(z)+ f_g(z)\sigma_{m}(z)\, \mu^2 \right]^2  + P_{shot},\nonumber\\
\ea
where we have defined
\ba
\sigma_{g}(z) &\equiv & b(z)\,G(z) \,\tilde{P}_0^{1/2}\nonumber\\
\sigma_{m}(z) &\equiv & G(z)\,\tilde{P}_0^{1/2},
\ea
The dimensionless power spectrum normalization constant $\tilde{P}_0$ is
just $P_0$ in Eq.(\ref{eq:P(k|z=0)}) in appropriate units:
\be
\label{eq:P0til}
\tilde{P}_0 \equiv \frac{P_0}{(\mbox{Mpc}/h)^3 (\mbox{Mpc})^{n_s}}
=\frac{\sigma_8^2}{I_0 \,h^{n_s}},
\ee
The second part of Eq.(\ref{eq:P0til}) is relevant if $\sigma_8$ is used to normalize the 
power spectrum. Note that
\be
I_0 \equiv  \int_0^\infty\rd \bar{k}\,  \frac{\bar{k}^{n_s+2}}{2\pi^2}\, 
T^2(\bar{k}\cdot h\mbox{Mpc}^{-1})\,
\left[\frac{3 j_1(8\bar{k})}{8\bar{k}}\right]^2,
\ee
where $\bar{k}\equiv k/[h\,\mbox{Mpc}^{-1}]$, 
and $j_1(kr)$ is spherical Bessel function. 
Note that $I_0=I_0(\omega_m, \omega_b, n_s, h)$. Since $k_{\parallel}$
and $k_\perp$ scale as $H(z)$ and $1/D_A(z)$ respectively, 
$\overline{P_g^{obs}(k)}$ in Eq.(\ref{eq:P(k)fg_scale1}) does not depend on $h$.

Eq.(\ref{eq:P(k)fg_scale1}) is analogous to the approach of
\cite{Song09}, who proposed the use of $f_g(z)\sigma_{8}(z)$ to
probe growth of large scale structure.
The difference is that Eq.(\ref{eq:P(k)fg_scale1}) uses $f_g(z)\sigma_{m}(z)
\equiv f_g(z)G(z) \tilde{P}_0^{1/2}$, which does {\it not} introduce an explicit 
dependence on $h$ (as in the case of using $f_g(z)\sigma_{8}(z)$).

The uncertainty in redshift measurements is included by multiplying
$P_g(k)$ with the damping factor, $e^{-k^2\mu^2\sigma_r^2}$, 
due to redshift uncertainties, with
\be
\sigma_r = \frac{\partial r}{\partial z}\, \sigma_z
\ee
where $r$ is the comoving distance. Note that the damping factor
should be held constant when taking derivatives of $P_g(k)$.

\subsection{$P(k)$ Method}
\label{sec:P(k)_method}

Including the nonlinear effects explicitly, we can write \citep{SE07}
\be
\frac{\partial P_g(k,\mu|z)}{\partial p_i}=
\frac{\partial P_g^{lin}(k,\mu|z)}{\partial p_i}\cdot
\exp\left(-\frac{1}{2}\, k^2\Sigma_{nl}^2\right).
\ee
The damping is applied to derivatives of $P_g(k)$, rather
than $P_g(k)$, to ensure that no information is extracted
from the damping itself. Eq.(\ref{eq:Fisher_full}) becomes
\ba
F_{ij}&=& V_{survey} \int_{-1}^{1}d\mu
\int_{k_{min}}^{k_{max}}
\frac{\partial \ln P_g^{lin}(k,\mu)}{\partial p_i}
\frac{\partial \ln P_g^{lin}(k,\mu)}{\partial p_j}\,\cdot\nonumber\\
& & \cdot \left[ \frac{n P_g^{lin}(k,\mu)}{nP_g^{lin}(k,\mu)+1}\right]^2
\, e^{-k^2\Sigma_{nl}^2} \, \frac{2\pi k^2 dk}{2\, (2\pi)^3}.
\label{eq:Fisher2_nl}
\ea
The linear galaxy power spectrum $P_g^{lin}(k,\mu|z)$ is given 
by Eq.(\ref{eq:P(k)fg_scale1}). The nonlinear damping scale 
\ba
\label{eq:NL}
\Sigma_{nl}^2&=&(1-\mu^2) \Sigma_{\perp}^2+\mu^2 \Sigma^2_{\parallel}
\nonumber\\
\Sigma_{\parallel}&=&\Sigma_{\perp} (1+f_g) \nonumber\\
\Sigma_{\perp}&= &12.4 \,h^{-1}{\rm Mpc}\, \left(\frac{\sigma_8}{0.9}\right)
\cdot 0.758\, G(z)\, p_{NL}\nonumber\\
&=& 8.355 \,h^{-1}{\rm Mpc}\, \left(\frac{\sigma_8}{0.8}\right)
\cdot G(z) \, p_{NL}.
\ea
The parameter $p_{NL}$ indicates the remaining level
of nonlinearity in the data; with $p_{NL}=0.5$ (50\% nonlinearity)
as the best case, and $p_{NL}=1$ (100\% nonlinearity)
as the worst case \citep{SE07}. For a fiducial
model based on WMAP3 results \citep{Spergel06}
($\Omega_m=0.24$, $h=0.73$, $\Omega_{\Lambda}=0.76$, $\Omega_k=0$,
$\Omega_b h^2=0.0223$, $\tau=0.09$, $n_s=0.95$, $T/S=0$),
$A_0=0.5817$, $P_{0.2}=2710\, \sigma_{8,g}^2$ \citep{SE07}.

In the $P(k)$ method, the full set of parameters that describe the observed $P_g(k)$ in each
redshift slice are: $\{\ln H(z_i)$, $\ln D_A(z_i)$, $\ln[f_g(z_i)\sigma_m(z_i)]$,
$\ln\sigma_{g}(z_i)$, $P_{shot}^i$; $\omega_m$, $\omega_b$, $n_s\}$,
where $\omega_m\equiv \Omega_m h^2$, and $\omega_b\equiv \Omega_b h^2$.
We marginalize over $\{\ln\sigma_{g}(z_i),P_{shot}^i\}$
in each redshift slice, to obtain a Fisher matrix for
$\{\ln H(z_i),\ln D_A(z_i),\ln[f_g(z_i)\sigma_m(z_i)]; \omega_m, \omega_b, n_s\}$.
This full Fisher matrix, or a smaller set marginalized over various parameters,
is projected into the standard set of cosmological parameters 
$\{w_0, w_a, \Omega_X, \Omega_k, \omega_m, \omega_b, n_s, \ln A_s\}$.
There are four different ways of utilizing the information from $P(k)$ (see Sec.\ref{sec:FoM}).

\subsection{BAO Only Method}
\label{sec:BAO_method}

The power of galaxy clustering as a dark energy probe was
first recognized via studies of baryon acoustic oscillations (BAO)
as a standard ruler \citep{BG03,SE03}.
The BAO only method essentially approximates $\partial P_g^{lin}(k,\mu)/\partial p_i$ with 
$\partial P_b^{lin}(k,\mu|z)/\partial p_i$ in the derivatives in Eq.(\ref{eq:Fisher2_nl}), with
the power spectrum that contains baryonic features, $P_b^{lin}(k,\mu)$, given by \citep{SE07}
\be
\label{eq:Pb_rev}
P_b^{lin}(k,\mu|z)= \sqrt{8\pi^2} A_0 \,P_g^{lin}(k_{0.2},\mu|z) 
\frac{\sin (x)}{x} \exp\left[ -(k \Sigma_s)^{1.4} \right],
\ee
where $P_g^{lin}(k,\mu|z)$ is the linear galaxy power spectrum, and
the Silk damping scale $\Sigma_s=8.38\,h^{-1}{\rm Mpc}$. We have defined
\ba
k_{0.2} &\equiv&  0.2\,h\,{\rm Mpc}^{-1}\\
x &\equiv&  \left(k^2_\perp s ^2_\perp +k^2_\parallel s^2_\parallel\right)^{1/2}
\ea
\noindent
Defining 
\ba
p_1 &=& \ln s_\perp^{-1} =\ln (D_A/s)\equiv \ln(x_h), \\
p_2 &=& \ln s_\parallel=\ln(s H)\equiv \ln(x_d),
\ea
substituting Eq.(\ref{eq:Pb_rev}) into Eq.(\ref{eq:Fisher2_nl}),
and making the approximation of $\cos^2 x \sim 1/2$, we find
\ba
&&\hspace{-0.1 in} F_{ij}\simeq V_{survey} A_0^2 \int_0^{1}\rd\mu 
\,f_i(\mu) \,f_j(\mu)\int_0^{k_{max}} \rd k \,k^2 \cdot \nonumber \\
& & \hskip 0.02in \cdot \left[\frac{ P^{lin}_m(k|z=0)}
{ P^{lin}_m(k_{0.2}|z=0)} +
\frac{1}{n P^{lin}_g(k_{0.2},\mu|z) \,e^{-k^2\mu^2\sigma_r^2}}\right]^{-2} 
\nonumber \\
& & \hskip 0.02in \cdot \exp\left[ -2(k \Sigma_s)^{1.4} -k^2\Sigma_{nl}^2\right],
\label{eq:Fisher_Wang}
\ea
where $P_g^{lin}(k_{0.2},\mu|z)$ is given by Eq.(\ref{eq:P(k)fg_scale1}) with $k=k_{0.2}$.

The functions $f_i(\mu)$ are given by
\ba
f_1(\mu)&=&\partial \ln x/\partial  p_1=\mu^2-1\\
f_2(\mu)&=&\partial \ln x/\partial  p_2=\mu^2.
\ea
The BAO only method gives $\{x_h(z), x_d(z)\}$ that are correlated at the level of
$\sim$41\% with each other, but are uncorrelated for different redshift slices
by construction.

\subsection{Assumptions and Priors}

We use the fiducial model adopted by the FoMSWG \citep{FoMSWG}, with
$\omega_m \equiv \Omega_m h^2=0.1326$, $\omega_b \equiv \Omega_b h^2=0.0227$, 
$h=0.719$, $\Omega_k=0$, $w=-1.0$, $n_s=0.963$, and $\sigma_8=0.798$.
No priors are used in deriving $\{x_h(z), x_d(z), f_g(z)\sigma_{m}(z)/s^4\}$,
which provide model-independent constraints on the cosmic expansion history
and the growth rate of cosmic large scale structure. These allow
the detection of dark energy evolution, and 
the differentiation between an unknown energy component and modified
gravity as the causes for the observed cosmic acceleration.

In order to derive dark energy figure of merit (FoM), as defined by
the DETF \citep{DETF}, we project our Fisher matrices into the standard 
set of dark energy and cosmological parameters:
$\{w_0, w_a, \Omega_X, \Omega_k, \omega_m, \omega_b, n_s, \ln A_s\}$.
To include Planck priors,\footnote{For a general and robust method for
including Planck priors, see \cite{Mukherjee08}.}
we convert the Planck Fisher matrix for 44 parameters 
(including 36 parameters that parametrize the dark energy equation of state in redshift 
bins) from the FoMSWG into a Planck Fisher matrix for this set of dark energy
and cosmological parameters.

We present all our results for StageIV+BOSS spectroscopic galaxy redshift surveys. 
The Stage IV galaxy redshift survey is assumed to cover 15,000 (deg)$^2$, with H$\alpha$ flux 
limit of $3\times 10^{-16}$ erg$\,$s$^{-1}$cm$^{-2}$, an efficiency of $e=0.50$, a redshift
range of $0.7<z<2.05$, and a redshift accuracy of $\sigma_z/(1+z)=0.001$. The galaxy number 
density is given by \cite{Geach10}, and the galaxy bias function is given by \cite{Orsi10}. 
This is similar to the baseline of the Euclid galaxy redshift survey \citep{RB}.
The BOSS survey is assumed to cover 10,000 (deg)$^2$, a redshift range of $0.1<z<0.7$, with 
a fixed galaxy number density of $n=3\times 10^{-4} h^3$Mpc$^{-3}$, and a fixed linear bias of
$b=1.7$.

\section{Results}

\subsection{Measurement of $H(z)$ and $D_A(z)$}

The $H(z)$ and $D_A(z)$ observables that correspond to the BAO scale are
\ba
x_h(z) & \equiv& H(z) s\nonumber\\
x_d(z) &\equiv & D_A(z)/s
\ea
where $s$ is the sound horizon scale at the drag epoch \citep{Hu96}. 

\begin{figure}
  \centering
  \includegraphics[trim = 0mm 0mm 60mm 0mm, width=0.9\columnwidth]{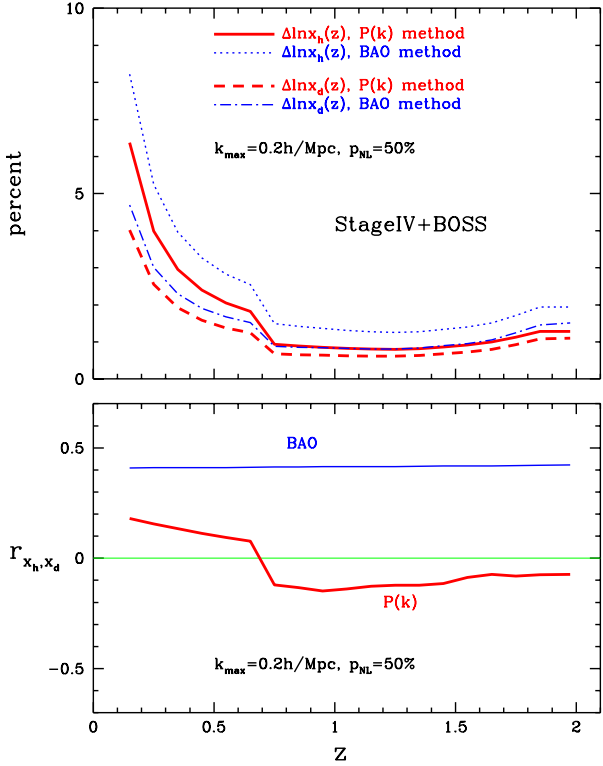}
  \caption{Precision of $x_h(z) \equiv H(z) s$ and $x_d(z) \equiv D_A(z)/s$ 
expected from StageIV+BOSS. 
The top panel shows the percentage errors on $x_h(z)$ and $x_d(z)$ 
per $\Delta z=0.1$ redshift bin, the bottom panel shows the 
normalized correlation coefficient between $x_h(z)$ and $x_d(z)$.}
  \label{fig:xhxd}
\end{figure}

Fig.\ref{fig:xhxd} shows the measurement precision of $x_h(z)$ and $x_d(z)$ for StageIV+BOSS. 
The top panel shows the percentage errors on $x_h(z)$ and $x_d(z)$, the bottom panel shows the 
normalized correlation coefficient between them. The thick solid and dashed lines
represent the measurement precision of $x_h(z)$ and $x_d(z)$ from the $P(k)$ method, marginalized over
all other parameters. The thin dotted and dot-dashed lines represent the measurement of 
$x_h(z)$ and $x_d(z)$ from the BAO only method.

Note that the $x_h(z)$ and $x_d(z)$ measured using $P(k)$ are only weakly correlated.
Since $x_h(z)$ and $x_d(z)$ represent independent degrees of freedom in a galaxy redshift
survey, they should not be strongly correlated. This is consistent with the findings of \cite{Chuang11}
from their analysis of SDSS LRG data.

The $x_h(z)$ and $x_d(z)$ from using the BAO method are correlated with a normalized
correlation coefficient of $r\sim 0.41$. They are positively correlated by construction:
In both $P(k)$ and BAO only methods, the Fisher matrix element for $\{x_h(z), x_d(z)\}$
from the same redshift slice is negative. In the $P(k)$ method, the $x_h(z)$ and $x_d(z)$
from different redshift slices are correlated through the cosmological parameters
$\{\omega_m, \omega_b, n_s\}$ that are measured using information from all the
redshift slices. When the cosmological parameters are marginalized over, the
dependence on these parameters remain as a weak correlation between $x_h(z)$ and $x_d(z)$.
In the BAO method, the $x_h(z)$ and $x_d(z)$ from different redshift slices are uncorrelated 
by construction. Inverting the 2$\times$2 Fisher matrix of $x_h(z)$ and $x_d(z)$ in each redshift
slice leads to positive and significant correlation between $x_h(z)$ and $x_d(z)$.

Finally, note that in using the $P(k)$ method to forecast dark energy constraints, 
two different methods have been used to account for nonlinear effects:\\
\noindent
(1) Setting $k_{max}=\pi/(2R)$, with $R$ given by requiring that 
$\sigma^2(R)$ is small (e.g., $\sigma^2(R)=0.25$), and imposing
a uniform upper limit cutoff, e.g., $k_{max}\leq 0.2h/\mbox{Mpc}$.
Note that in this case $p_{NL}=0$ in Eq.(\ref{eq:NL}); the nonlinear effects are minimized
by imposing a minimum length scale that increases at lower redshift.\\
\noindent
(2) Setting $k_{max}$ to a fixed value, and account for nonlinear
effects through the exponential damping term in Eq.(\ref{eq:NL}),
e.g., $p_{NL}=0.5$.

With a suitable choice of $\sigma^2(R)$ in (1) and $k_{max}$ in (2),
these two methods of accounting for nonlinear effects give the same
DETF dark energy FoM. 
For StageIV+BOSS, for the $P(k)$ only method (no priors and marginalizing over growth information),
(1) with $\sigma^2(R)=0.28$ and $k_{max} \leq 0.2h/\mbox{Mpc}$ gives FoM$=$49.9,
while (2) with $p_{NL}=0.5$ and $k_{max}= 0.2h/\mbox{Mpc}$ gives FoM$=$49.6.
These two cases give very similar uncertainties on $x_h(z)$ and $x_d(z)$.

Since these two nonlinear cutoff methods are very similar, we have chosen to use
cutoff method (2) in the rest of this paper, since it is smooth with $k$, 
and is the approach used in the BAO only method.

\subsection{Growth Rate Measurements}
\label{sec:growth}

\begin{figure}
  \centering
  \includegraphics[trim = 0mm 0mm 60mm 0mm, width=0.9\columnwidth]{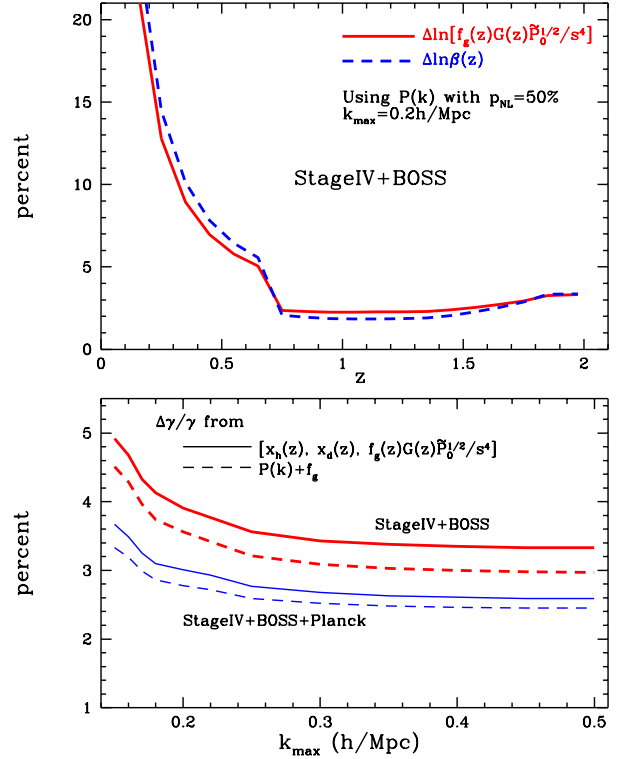}
  \caption{Top: uncertainties on $f_g(z)G(z) \tilde{P}_0^{1/2}/s^4$ and $\beta(z)$ for StageIV+BOSS
per $\Delta z=0.1$ redshift bin. 
Bottom: uncertainties on the growth rate powerlaw index $\gamma$ for StageIV+BOSS, with and
without Planck priors.}
  \label{fig:fgsmb}
\end{figure}

\cite{Song09} showed that assuming a linear bias between galaxy and
matter distributions, we can use $f_g(z)\sigma_{8}(z)$ to probe gravity without
additional assumptions. We use a similar approach, but use
$f_g(z)\sigma_{m}(z)\equiv f_g(z)G(z) \tilde{P}_0^{1/2}$ to avoid
introducing an explicit dependence on $h$ through $\sigma_8$.
We find that the $f_g(z)\sigma_{m}(z)$ measurements from
the $P(k)$ method are highly correlated with the $\omega_m$ measurement, which makes
the uncertainties on $f_g(z)\sigma_{m}(z)$ much larger than that of
$\beta(z)$. Fortunately, we are able to find a scaled measurement of
$f_g(z)\sigma_{m}(z)$, 
\be
\label{eq:fg_scaled}
\overline{f_g(z)\sigma_{m}(z)} \equiv \frac{f_g(z)\sigma_{m}(z)}{s^4}
\equiv \frac{f_g(z)G(z) \tilde{P}_0^{1/2}}{s^4},
\ee
that is nearly uncorrelated with $\omega_m$, and has an uncertainty that approaches that
of $\beta(z)$, see top panel of Fig.\ref{fig:fgsmb}. 
The precision of both $f_g(z)\sigma_{m}(z)/s^4$ and $\beta(z)$ are insensitive to the 
choice of $k_{max}$.

To make sense of the scaling in Eq.(\ref{eq:fg_scaled}), note that the
observed power spectrum depends on $\omega_m$ only through $s$ and $T(k)$,
as follows (see Eq.[\ref{eq:P(k)fg_scale1}]) for $P_{shot}=0$:
\ba
P_g^{obs} &\propto &\frac{x_h(z)}{x_d^2(z)}\cdot \frac{1}{s^3}\cdot
\left[ \sigma_{g}(z)+ f_g(z)\sigma_{m}(z)\, \mu^2 \right]^2
\cdot \frac{(ks)^{n_s}}{s^{n_s}} \cdot T^2(k) \nonumber\\
&\propto & 
\left[ \sigma_{g}(z)+ f_g(z)\sigma_{m}(z)\, \mu^2 \right]^2
\, T^2(k) \, s^{-(n_s+3)}.
\ea
Note that at the peak of $P_m(k|z=0)=P_0 k^{n_s}T^2(k)$, $k=k_p$,
\be
\label{eq:kp}
\frac{n_s}{\tilde{k}_p}+ 2 \left.\frac{\partial \ln T(k)}{\partial \tilde{k}}\right|_{\tilde{k}_p}=0,
\ee
where $\tilde{k} \equiv k/$Mpc$^{-1}$. $T(k)$ depends only on
\be
q \equiv \frac{k}{h\, \mbox{Mpc}^{-1}}\, \Theta^2_{2.7}/\Gamma
\simeq \frac{\tilde{k}}{\omega_m},
\ee
where $\Theta_{2.7}\equiv T_{CMB}/2.7$K, $\Gamma\simeq \Omega_m h$ \citep{EH98}.
Thus at $k=k_p$,
\be
\label{eq:dlnT}
\frac{\partial \ln T(k)}{\partial \tilde{k}} \simeq \frac{1}{\omega_m}\frac{\rd \ln T}{\rd q},
\ee
and we find
\ba
\left.\frac{\partial \ln T}{\partial \omega_m}\right|_{k_p}
&\simeq & \left.\frac{\rd \ln T}{\rd q} \cdot \frac{\partial q}{\partial \omega_m}\right|_{\tilde{k}_p}\nonumber\\
& =& -\frac{\tilde{k}_p}{\omega_m^2}\cdot \frac{\rd \ln T}{\rd q}\nonumber\\
&=& \frac{n_s}{2\omega_m},
\ea
where we have used Eqs.(\ref{eq:kp}) and (\ref{eq:dlnT}).
Using the approximate formula for $s$ from \cite{EH98},
\be
s \simeq \frac{44.5 \ln (9.83/\omega_m)}{\sqrt{1+10 \omega_b^{3/4}}}\,\mbox{Mpc}
\ee
we find that at the peak of $P(k)$,
\ba
\left.\frac{\partial \ln T}{\partial \omega_m}\right|_{k_p} &\simeq&
- \frac{n_s}{2} \, \ln(9.83/\omega_m)\, \frac{\partial \ln s}{\partial \omega_m}\\
&\simeq& -2 \, \frac{\partial \ln s}{\partial \omega_m}.
\label{eq:dTdomhh-2}
\ea
We have assumed $\omega_m$ and $n_s$ close to our fiducial values of $\omega_m=0.1326$
and $n_s=0.963$ in obtaining Eq.(\ref{eq:dTdomhh-2}).

If we define the scaled parameters
\ba
\label{eq:fg_scale2}
&&\overline{\sigma_{g}(z)}\equiv\frac{\sigma_{g}(z)}{s^4}
=\frac{b(z)G(z)\tilde{P}_0^{1/2}}{s^4 },\\
&&\overline{f_g(z)\sigma_{m}(z)} \equiv \frac{f_g(z)\sigma_{m}(z)}{s^4}
=\frac{f_g(z)G(z)\tilde{P}_0^{1/2}}{s^4},\nonumber
\ea
we find
\be
P_g^{obs} \propto \frac{x_h(z)}{x_d^2(z)}\cdot 
\left[ \overline{\sigma_{g}(z)}+ \overline{f_g(z)\sigma_{m}(z)}\, \mu^2 \right]^2
\cdot (ks)^{n_s} s^{5-n_s} T^2(k).
\ee
In the new set of parameters, $\{x_h(z_i)$, $x_d(z_i)$, $\overline{f_g(z_i)\sigma_{m}(z_i)}$,
$\overline{{\sigma}_{g}(z_i)}$, $P_{shot}^i$; $\omega_m$, $\omega_b$, $n_s\}$,
the dependence of $P_g^{obs}$ on $\omega_m$ only comes through
the combination of $s^{5-n_s} T^2(k)\simeq [s^2 T(k)]^2$, which is
only very weakly dependent on $\omega_m$ (see Eq.[\ref{eq:dTdomhh-2}]).
Thus the dependence of $P_g^{obs}$ on $\omega_m$ is effectively removed or absorbed via
the scaling of parameters in Eq.(\ref{eq:fg_scale2}), leading
to measurements on $\overline{f_g(z)\sigma_{m}(z)}$ that are essentially
uncorrelated with $\omega_m$, and greatly improved in precision over
that of $f_g(z)\sigma_{m}(z)$.
This is as expected, since the measurements of $f_g(z)\sigma_{m}(z)$
are strongly correlated with that of $\omega_m$ (i.e., $P(k)$ shape).

The bottom panel of Fig.\ref{fig:fgsmb} shows the uncertainties on the growth rate powerlaw 
index $\gamma$ for StageIV+BOSS, with and without Planck priors. Note that $\gamma$ is defined by
parametrizing the growth rate as a powerlaw \citep{WangStein98,Lue2004},
\be
f_g(z)=\left[ \Omega_m(a)\right]^{\gamma},
\ee
where $\Omega_m(a)=8\pi G\rho_m(a)/(3H^2)$.
The solid lines in the bottom panel of Fig.\ref{fig:fgsmb} show the precision on
$\gamma$ using only the $\{x_h(z), x_d(z), f_g(z)\sigma_{m}(z)/s^4\}$ measured from
$P(k)$ and marginalized over all other parameters. The dashed lines show the
precision on $\gamma$ when the full $P(k)$ is used, including the growth information
(i.e., the ''$P(k)+f_g$'' method).

\subsection{Dark Energy Figure of Merit}
\label{sec:FoM}

To calculate the DETF dark energy FoM \citep{DETF}, FoM$=1/\sqrt{\mbox{det}[{\rm Cov}(w_0,w_a)]}$ \citep{Wang08b}, 
we need to project our large set of measured parameters into the standard set of
$\{w_0, w_a, \Omega_X, \Omega_k, \omega_m, \omega_b, n_s, \ln A_s\}$.

The BAO only method gives measurement of $\{x_h(z_i), x_d(z_i)\}$ from each redshift slice.
The Fisher matrix for these measurements are then projected into the Fisher matrix
for $\{w_0, w_a, \Omega_X, \Omega_k, \omega_m, \omega_b\}$. Because the dependence
on $\{\omega_m, \omega_b\}$ only comes through $s$, $\omega_m$ and $\omega_b$ are perfectly
degenerate if no priors are added. This can be shown explicitly by computing the
submatrix for $\{\omega_m, \omega_b\}$ in the Fisher matrix, which is proportional to
\ba
\left(
\begin{array}{cc}
\left(\frac{\partial \ln x_h}{\partial \omega_m}\right)^2 & 
\left(\frac{\partial \ln x_h}{\partial \omega_m}\right)\left(\frac{\partial \ln x_h}{\partial \omega_b}\right)\\
\left(\frac{\partial \ln x_h}{\partial \omega_m}\right)\left(\frac{\partial \ln x_h}{\partial \omega_b}\right)&
\left(\frac{\partial \ln x_h}{\partial \omega_b}\right)^2
\end{array}
\right)
\ea
the determinant of this submatrix is zero, thus the determinant of the entire
Fisher matrix for $\{w_0, w_a, \Omega_X, \Omega_k, \omega_m, \omega_b\}$ is zero. 
It can be shown that the combination determined by the BAO only method is
\ba
\tilde{\omega}_m &\equiv& \omega_m+\omega_b\, \left(\frac{\partial \ln x_h}{\partial \omega_b}\right)
\left(\frac{\partial \ln x_h}{\partial \omega_m}\right)^{-1}\nonumber\\
&=&\omega_m+\omega_b\, \left(\frac{\partial \ln s}{\partial \omega_b}\right)
\left(\frac{1}{2\omega_m}+ \frac{\partial \ln s}{\partial \omega_m}\right)^{-1}
\ea
It can be shown explicitly that the Fisher matrix for 
$\{w_0, w_a, \Omega_X, \Omega_k, \tilde{\omega}_m\}$ is exactly the same 
as the $\{w_0, w_a, \Omega_X, \Omega_k, \omega_m\}$ submatrix of the original
Fisher matrix for $\{w_0, w_a, \Omega_X, \Omega_k, \omega_m, \omega_b\}$.
Thus to compute the FoM for BAO only, one only needs to drop the Fisher matrix
elements for $\omega_b$, then invert the resultant Fisher matrix to obtain the covariance
matrix. Note that the Fisher matrix for 
$\{w_0, w_a, \Omega_X, \Omega_k, \omega_m, \omega_b\}$ should be used
when combining with Planck priors.

There are four different ways that we can extract dark energy information 
from the $P(k)$ method (in the order of increasing information content):\\
\noindent
(1) $\{x_h(z), x_d(z)\}$ from $P(k)$:
Project the Fisher matrix for $\{ \ln H(z_i)$, $\ln D_A(z_i)$, $\ln[f_g(z_i)\sigma_{m}(z_i)]$; 
$\omega_m$, $\omega_b$, $n_s\}$ into that of $\{\ln x_h(z_i)$, $\ln x_d(z_i)$, 
$\ln[f_g(z_i)\sigma_{m}(z_i)/s^4]$; $\omega_m$, $\omega_b$, $n_s\}$
(see Sec.\ref{sec:growth}),
then marginalize over $\{\ln[f_g(z_i)\sigma_{m}(z_i)/s^4]; \omega_m, \omega_b, n_s\}$,
and project the Fisher matrix for $\{\ln x_h(z_i), \ln x_d(z_i)\}$ into the
standard set of cosmological parameters.\\
\noindent
(2) $\{x_h(z), x_d(z), f_g(z)\sigma_{m}(z)/s^4\}$ from ${\boldmath P(k)}$:
Project the Fisher matrix for $\{\ln H(z_i)$, $\ln D_A(z_i)$, $\ln[f_g(z_i)\sigma_{m}(z_i)]$; 
$\omega_m$, $\omega_b$, $n_s\}$ into that of $\{\ln x_h(z_i)$, $\ln x_d(z_i)$, 
$\ln[f_g(z_i)\sigma_{m}(z_i)/s^4]$; $\omega_m$, $\omega_b$, $n_s\}$
(see Sec.\ref{sec:growth}), then marginalize over $\{\omega_m, \omega_b, n_s\}$,
and project the Fisher matrix for $\{\ln x_h(z_i), \ln x_d(z_i), \ln[f_g(z_i)\sigma_{m}(z_i)/s^4]\}$ into the standard set of cosmological parameters.\\
\noindent
(3) ${\boldmath P(k)}$ marginalized over ${\boldmath f_g}$:
Marginalize over $\ln[f_g(z_i)\sigma_{m}(z_i)]$ to obtain the Fisher matrix for
$\{\ln H(z_i),\ln D_A(z_i); \omega_m, \omega_b, n_s\}$, and project it into the 
standard set of cosmological parameters.\\
\noindent
(4) ${\boldmath P(k)+f_g}$:
Project the Fisher matrix for $\{\ln H(z_i)$, $\ln D_A(z_i)$, $\ln[f_g(z_i)\sigma_{m}(z_i)]$;
$\omega_m$, $\omega_b$, $n_s\}$ into the standard set of cosmological parameters.

\begin{figure}
  \centering
  \includegraphics[trim = 0mm 0mm 60mm 0mm, width=0.9\columnwidth]{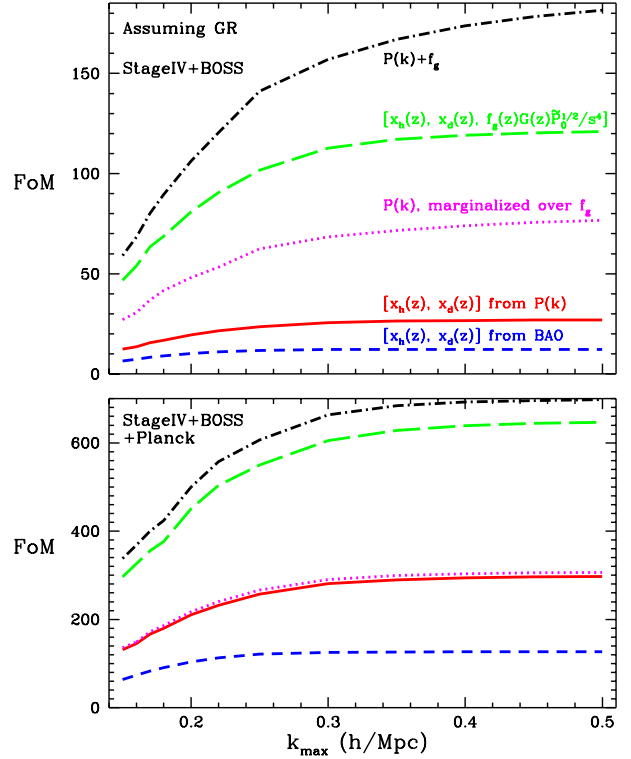}
  \caption{Dark energy FoM for StageIV+BOSS, without (top panel) and with (bottom panel) Planck priors,
as a function of the nonlinear cutoff $k_{max}$. }
  \label{fig:FoM}
\end{figure}

Fig.\ref{fig:FoM} shows the DETF dark energy FoM for StageIV+BOSS, without (top panel) and 
with (bottom panel) Planck priors, as a function of the nonlinear cutoff $k_{max}$. 
The four methods of using $P(k)$ described above, as well as the BAO only method, are shown. 
Note that the FoMs from the three most conservative methods, BAO only, $\{x_h(z), x_d(z)\}$ 
from $P(k)$, and $\{x_h(z), x_d(z), f_g(z)\sigma_{m}(z)/s^4\}$ from $P(k)$, are insensitive to the
increase of $k_{max}$ for $k_{max} \ga 0.3h/$Mpc. We have not included
the nonlinearity in the RSD due to peculiar velocities here for simplicity.
Adding a peculiar velocity of 300$\,$km/s is equivalent to adding $0.001$
in quadrature to the redshift dispersion $\sigma_z=0.001(1+z)$; this has a
negligible effect on the FoM for StageIV+BOSS, since the FoM is most sensitive to
assumptions about the Stage IV survey, which is at $z\geq 0.7$.

The most conservative of the $P(k)$ approaches, using $\{x_h(z), x_d(z)\}$ measured from
$P(k)$ and marginalized over all other parameters, gives a dark energy FoM about
a factor of two larger than that of the BAO only method, with or without Planck priors.
This provides a robust conservative method to go beyond BAO only in extracting 
dark energy information from galaxy clustering data.

It is interesting to note that $\{x_h(z), x_d(z)\}$ from $P(k)$ (solid line) gives similar 
dark energy FoM to that of the full $P(k)$ marginalized over growth information (dotted line), 
when Planck priors are included. Similarly, $\{x_h(z), x_d(z), f_g(z)\sigma_{m}(z)/s^4\}$ gives 
dark energy FoM close to that of the full $P(k)$ with growth information included, when
Planck priors are added.

\subsection{Comparison With Previous Work}

This work has the most overlap with \cite{Wang_etal10}, which explored
the optimization of a space-based galaxy redshift survey.
The differences of this work from \cite{Wang_etal10} are:\\
(1) This work presents a {\it new} conservative approach to extract dark energy
constraints from galaxy clustering data: the use of {\it only} the $H(z)s$ and $D_A(z)/s$
measurements from the observed galaxy power spectrum, $P_g^{obs}(k)$, to probe dark energy.
This bridges the methods using $P_g^{obs}(k)$ and the BAO method (which uses
$H(z)s$ and $D_A(z)/s$ measurements from fitting the BAO peaks).\\
(2) This work presents a {\it new} combination of growth information,
$f_g(z) G(z)/s^4$, that can be measured nearly as precisely as the
linear redshift-space distortion parameter $\beta$ (see Fig.2), but can
be used to probe the growth history of cosmic large scale structure
without assuming a bias model.
\cite{Wang_etal10} did not study growth constraints explicitly;
they either marginalized over the growth rate
information, or assumed that gravity is described by general relativity.\\
(3) This work focuses on model-independent constraints of dark energy and gravity
in terms of $H(z)s$, $D_A(z)/s$, and $f_g(z) G(z)/s^4$ measured in $\Delta z=0.1$ redshift bins.
\cite{Wang_etal10} focuses on the conventional dark energy model
with dark energy equation of state given by $w_X(z)=w_0+w_a (1-a)$ \citep{Chev01}, and 
dark energy density function $X(z)=\rho_X(z)/\rho_X(0)$ parametrized by its
value at $z=2/3$, 4/3, and 2.

The methodology developed in this work differs from what is currently used
in analyzing galaxy clustering data.  This work proposes the simultaneous
measurement of $H(z)s$, $D_A(z)/s$, and $f_g(z)G(z)/s^4$ from galaxy clustering
data without imposing any priors.
Because of the limited volume probed by current data, no simultaneous measurements
of $H(z)$, $D_A(z)$, and $f_g(z)$ have been made without imposing strong priors
on cosmological parameters. The first simultaneous measurements of
$H(z)s$ and $D_A(z)/s$ were made by \cite{Chuang11} at $z_{eff}=0.35$ using SDSS DR7 LRG data; 
they marginalized over growth information.
\cite{Blake11} measured $f_g(z)\sigma_8(z)$ at several redshifts
while fixing the background cosmology. Most recently, 
\cite{Reid12} published the first simultaneous measurement of
$D_A(z)H(z)$, $D_A(z)^2/H(z)$, and $f_g(z)\sigma_8(z)$ at $z_{eff}=0.57$
using BOSS data, assuming WMAP7 priors.

This work presents forecasts of the precision of the most general measurements of 
cosmic expansion history (via $H(z)$ and $D_A(z)$) and gravity (via
$f_g(z)G(z)$) that can be made from a Stage IV galaxy redshift survey.
These will be the most model-independent results on probing dark energy
and probing gravity from such a survey.

\section{Summary and Discussion}

We have examined how the constraints on the scaled expansion history of the universe, 
$x_h(z)=H(z)s$, and the scaled angular diameter distance, $x_d(z)=D_A(z)/s$, depend on the methods 
used to analyze the galaxy clustering data. We find that using the observed galaxy power 
spectrum, $P_g^{obs}(k)$, $x_h(z)$ and $x_d(z)$ are measured more accurately and are significantly 
less correlated with each other, compared to using only the information from the baryon 
acoustic oscillations (BAO) in $P_g^{obs}(k)$ (see Fig.1). Using the $\{x_h(z), x_d(z)\}$ from $P_g^{obs}(k)$ 
gives a DETF dark energy FoM approximately a factor of two larger than using the 
$\{x_h(z), x_d(z)\}$ from BAO only (see Fig.3); this provides a robust conservative method to go beyond 
BAO only in extracting dark energy information from galaxy clustering data.
This is encouraging since \cite{Chuang11} found that $\{x_h(z), x_d(z)\}$ from SDSS
galaxy clustering data are not sensitive to systematic uncertainties.

Furthermore, we find that if the redshift-space distortion information 
contained in $P_g^{obs}(k)$ is used, we can measure $\{x_h(z), x_d(z), f_g(z)G(z)\tilde{P}_0^{1/2}/s^4\}$
with high precision from a Stage IV galaxy redshift survey with $0.7<z<2$ over
15,000 (deg)$^2$ (see Figs.1 and 2), where $f_g(z)$ and $G(z)$ are linear growth rate and
growth factor of large scale structure respectively, and $\tilde{P}_0$ denotes the dimensionless
normalization of $P_m^{lin}(k|z=0)$ . Adding $f_g(z)G(z)\tilde{P}_0^{1/2}/s^4$ to 
$\{x_h(z), x_d(z)\}$ significantly boosts the dark energy FoM, compared to using $\{x_h(z), x_d(z)\}$ only, or
using $P_g^{obs}(k)$ marginalized over the growth information, assuming that gravity is not
modified (see Fig.3). Alternatively, $f_g(z)G(z)\tilde{P}_0^{1/2}/s^4$ provides a powerful test 
of gravity, as dark energy and modified gravity models that give identical $H(z)$
likely give different $f_g(z)$ \citep{Wang08a}. Measuring
$\{x_h(z), x_d(z), f_g(z)G(z)\tilde{P}_0^{1/2}/s^4\}$ simultaneously
allows us to probe gravity without fixing the background
cosmological model. We will be adopting this approach to analyze simulated and real
galaxy redshift catalogs in future work.

We have developed a conservative approach to analyzing galaxy clustering
data that should be insensitive to systematic uncertainties, if only
data on quasi-linear scales are used ($k_{max}\la 0.2h/$Mpc). 
Since the dark energy FoM (see Fig.3) and the gravity 
constraints (see lower panel of Fig.2) are insensitive to the inclusion of 
smaller scale information at $k_{max}\ga 0.3h/$Mpc, our results are likely robust indicators of
how well a Stage IV galaxy redshift survey can probe dark energy and
constrain gravity.

In analyzing real data, the systematic 
effects (bias between luminous matter and matter distributions, nonlinear effects, 
and redshift-space distortions)\footnote{See, e.g., \cite{BG03,SE03}. For reviews,
see \cite{Wang_book} and \cite{Weinberg12}.} will ultimately need to be modeled 
in detail and reduced where possible (see, e.g., \cite{Percival10,Blake11,Padmanabhan12}). 
This will require cosmological N-body simulations that include galaxies,
either by incorporating physical models of galaxy formation (see, e.g., \cite{Baugh06,Angulo08}), 
or using halo occupation distributions (HOD) measured from the largest available data sets
(see, e.g., \cite{Zheng09}, and http://lss.phy.vanderbilt.edu/lasdamas/overview.html).
We can expect a Stage IV galaxy redshift survey to play a critical role 
in advancing our understanding of cosmic acceleration within the next decade.

\section*{Acknowledgments}

I am grateful to Chia-Hsun Chuang, and especially Will Percival for very useful discussions.
This work was supported in part by DOE grant DE-FG02-04ER41305.

\setlength{\bibhang}{2.0em}

\label{lastpage}

\end{document}